\documentclass[twocolumn,showpacs,superscriptaddress,preprintnumbers,amsmath,
amssymb,prb]{revtex4}

\usepackage{graphicx}
\usepackage{dcolumn}
\usepackage{bm}
\usepackage{hyperref}
\usepackage{setspace}

\begin{document}

\title{First principles study of the spin state transitions in GdBaCo$_2$O$_{5.5}$
}

\author{V. Pardo}
 \email{vpardo@usc.es}
\affiliation{
Departamento de F\'{\i}sica Aplicada, Facultad de F\'{\i}sica, Universidad
de Santiago de Compostela, E-15782 Campus Sur s/n, Santiago de Compostela,
Spain
}
\affiliation{
Instituto de Investigaciones Tecnol\'ogicas, Universidad de Santiago de
Compostela, E-15782, Santiago de Compostela, Spain
}

\author{D. Baldomir}
\affiliation{
Departamento de F\'{\i}sica Aplicada, Facultad de F\'{\i}sica, Universidad
de Santiago de Compostela, E-15782 Campus Sur s/n, Santiago de Compostela,
Spain
}
\affiliation{
Instituto de Investigaciones Tecnol\'ogicas, Universidad de Santiago de
Compostela, E-15782, Santiago de Compostela, Spain
}

\begin{abstract}

Electronic structure calculations were carried out on the compound
GdBaCo$_2$O$_{5.5}$. 
The electronic structure variation with a change in the
spin state of the Co$^{3+}$ ion in an octahedral environment has
been studied. All the insulating phases are described and possible scenarios
for the metallic ones are presented.
Orbital ordering is shown to take place and the electronic structure leading
to it is determined.
The Ising-like anisotropy shown experimentally
can be predicted.
Also, big unquenched orbital angular
momenta are calculated and their origin is described.

\end{abstract}


\maketitle

\newpage

\section{Introduction}

Co oxides have been the objective of deep research during the last few
years. As any correlated oxide, they present an interplay between the electronic structure,
magnetic properties and geometric structure that makes them very
interesting from both an experimental and a theoretical point of view.
They have received a strong attention very recently, mainly after the
discovery of both superconductivity\cite{co_sc} and
magnetoresistance\cite{co_cmr} in them because of the vast technological applications of those phenomena, but
also because of the magnetic transitions accompanied of
structural transitions, that are not yet fully understood.
GdBaCo$_2$O$_{5.5}$ presents magnetic transitions at about 75 and 220 K and a
metal-insulator transition at some 360 K.\cite{magnprops} The
magnetization measurements at the different phases have not been
conclusively ascribed to a particular spin state and orbital
configuration. There are no measurements available of the orbital
magnetic moments of the Co ions in the different phases and spin states.
This rich variety of phases including spin state transitions occurs in Co$^{3+}$ compounds, where the low-spin state is not stable
at all temperatures because of the small magnitude of the crystal field,
comparable to the intra-atomic exchange energy. This is the case of GdBaCo$_2$O$_{5.5}$, that also
presents giant magnetoresistance\cite{co_cmr} associated with a
metal-insulator transition. Very recently, a spin blockade phenomenon has
been found to explain its conduction
properties.\cite{blockade,taskin_blockade} Some work has also been carried
out from 
a theoretical standpoint.\cite{hf_gdbacoo,hf_prb,wu_prb2,khalyavin} 
The goal of this paper is to make use of electronic structure
calculations in order to explain the magnetic and
electronic structure of the compound and their variations with respect to the magnetic and spin state
transitions that occur in it.
For doing so, in section \ref{strsec} we will describe the structure of
the material and the method of calculation and in section \ref{elecsec} we will
present our calculations of the magnetic and electronic structure, dealing
with the spin state, magnetic and metal-insulator transitions, and also with the orbital
ordering phenomenon
relating all of these to the
macroscopic observations
establishing a consistent picture for the observed
properties of the material.

\section{Structure and computational details}\label{strsec}

\begin{figure}
\includegraphics[width=\columnwidth]{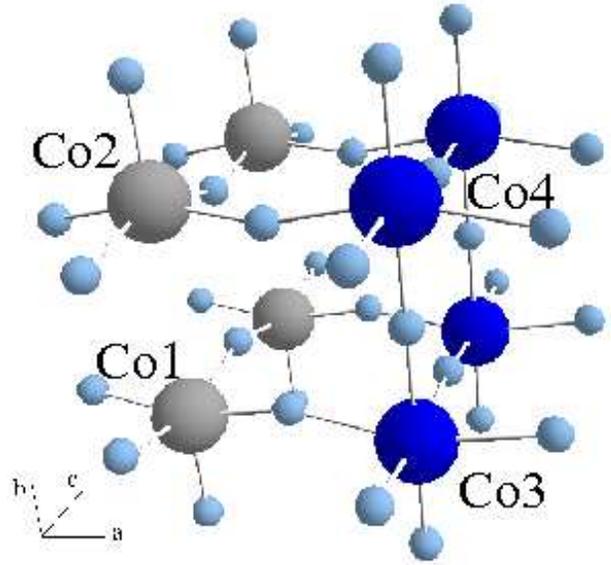}
\caption{(Color online) Structure of GdBaCo$_2$O$_{5.5}$. Grey atoms (Co1
and Co2) are Co$_{pyr}$,
dark atoms (Co3 and Co4) are Co$_{oct}$ and the small light ones are the oxygen atoms.
Observe the oxygen vacants along the c-axis leading to a square
pyramidal environment. 
}\label{strfig}
\end{figure}

GdBaCo$_2$O$_{5.5}$ presents 
two crystallographically different surroundings
for the two inequivalent Co ions in the unit cell. One of them is in a
square pyramidal (Co$_{pyr}$) and the other in an
octahedral environment (Co$_{oct}$). 
There is one oxygen vacant per unit cell, that produces the square
pyramidal environment for the Co in the crystallographic position 2r.
The octahedron surrounding Co$_{oct}$ is elongated along the $a$-axis (see Fig.
\ref{strfig} for the
naming of the crystallographic directions). 
The unit cell is orthorhombic with space group
Pmmm. The value of the lattice parameters and atomic positions were taken
from Respaud \emph{et al.}\cite{frontera_prb} ($a$= 3.87738 \AA, b=
7.53487 \AA\  and c= 7.8269 \AA), but the oxygen positions were recalculated
performing a full structural relaxation.
It is well established that the oxygen vacants are in the
crystallographic position 1g (1/2, 1/2, 1/2).\cite{frontera_prb} For our
calculations we broke the symmetry along the b-axis, so that 4 
inequivalent Co atoms enter the cell (let us call Co1 and Co2 the two
inequivalent atoms in the pyramidal position and Co3 and
Co4 the two inequivalent atoms in the octahedral environment, as can be
seen in Fig. \ref{strfig}). Further symmetry breakings were produced along
the c- and $a$-axis to have 8 Co atoms in the supercell: the former case
with the intention of studying possible spin state orderings in the $yz$
plane and the
latter for analyzing the magnetic couplings between Co$_{pyr}$ layers
mediated by a plane of Co$_{oct}$ atoms.

\begin{table}[h]
\caption{Crystallographic positions at T= 0 that result from our structure
optimization.}\label{strtab}
\begin{tabular}{|c|c|c|}
\hline
Atom & Crystallographic & Coordinates\\
& position &\\
\hline
\hline
Ba & 2$\sigma$ & (0,0,0.2503)\\
Gd & 2p & (0,0.5,0.2292)\\
Co$_{pyr}$ & 2r & (0.5,0.2563,0.5)\\
Co$_{oct}$ & 2q & (0.5,0.2487,0)\\
O1 & 1a & (0.5,0,0)\\
O2 & 1e & (0.5,0,0.5)\\
O3 & 1c & (0.5,0.5,0)\\
O4 & 2s & (0,0.2731,0)\\
O5 & 2t & (0,0.3102,0.5)\\
O6 & 4u & (0.5,0.2947,0.2633)\\
\hline
\end{tabular}
\end{table}

We present here full-potential, all-electron, electronic structure
calculations based on the density functional theory (DFT) utilizing the APW+lo
method \cite{sjo} performed with the WIEN2k software.\cite{wien,wien2k} For
the structure minimization, the results of which are summarized in Table
\ref{strtab}, we used the GGA (generalized gradient
approximation) in the Perdew-Burke-Ernzerhoff (PBE) scheme.\cite{gga} 
The geometry optimization was carried out minimizing the forces in the
atoms and the total energy of the system. These results are independent of the
magnetic ordering, forces are minimized both in the case of ferromagnetic (FM) or antiferromagnetic
(AF) ordering. Also, a force minimization procedure was carried out in a
supercell containing 8 inequivalent Co atoms. Minimal changes of the
atomic positions
compared to those reported in Table \ref{strtab} turn out as a result.
For the electronic structure calculations, we included the strong
correlation effects by means of the LDA+U scheme\cite{sic} in the
so-called ``fully-localized limit".\cite{fll} The values of U chosen were
within the range of 3 to 6 eV for all the spin states (U depends on the spin
state of the atom) but all the results presented here are fully consistent for
U within that range (the quantitative results of energy differences depend on the choice
of U by some 25\% for values of U within that range, the values presented
will be given for U= 4.5 eV).
Spin-orbit effects have been
introduced in a second variational way using the scalar relativistic
approximation.\cite{singh} 
The parameters of our calculations depend on the type of calculation but
for any of them we converged with respect to the k-mesh and to
R$_{mt}$K$_{max}$, up to 1000 k-points (256 in the irreducible Brillouin
zone) and up to R$_{mt}$K$_{max}$= 7. Local orbitals were added for a bigger
flexibility in dealing with the semicore states. Muffin-tin radii chosen
were the following: 2.3 a.u. for both Ba and Gd, 1.90 a.u. for Co and 1.59
a.u. for O.

\section{Electronic and magnetic structure}\label{elecsec}

From a purely ionic point of view, we expect both Co$^{3+}$ ions
in a d$^6$ configuration, that in an octahedral environment could lead to
either a low spin (LS) state (t$_{2g}^{6}$e$_{g}^{0}$; nonmagnetic S=0), an
intermediate spin (IS) state 
(t$_{2g}^{5}$e$_{g}^{1}$; S=1) or even a high spin (HS) state 
(t$_{2g}^{4}$e$_{g}^{2}$; S=2). Spin state transitions play an important
role in the temperature evolution of the electronic, structural and
magnetic properties of
transition metal oxides, specially in the case of Co$^{3+}$ compounds,
such as the famous example of
LaCoO$_3$.\cite{lacoo} In the case of GdBaCo$_2$O$_{5.5}$, it is not yet clearly understood
the relationship between a spin state transition and the transition from
AF to FM behavior\cite{kim}
(neither with the metal-insulator transition).\cite{frontera_jssc}
We have studied in this paper
how the electronic structure evolves with spin states and our calculations
always relax to a scenario with 
Co$_{pyr}$ in the IS state, not being possible the stabilization of a
different spin state for that atom. Usually, the pryramidal sites of
GdBaCo$_2$O$_{5.5}$ have
been ascribed to the IS state,\cite{maignan_jssc, respaud_prb} even crystal field arguments
confirm it as being very stable.\cite{magnprops} Very recently, a spin
blockade phenomenon was found to explain the conduction properties of the
material,\cite{blockade} being its origin explainable assuming an IS state
in Co$_{pyr}$ atoms.\cite{taskin_blockade} However, some works evidence that a HS state is
likely to occur for a Co$^{3+}$ ion in a square pyramidal environment in
some cases, as has been evidenced by neutron diffraction,\cite{neutron1,
neutron2} spectroscopy measurements,\cite{spectroscopy} and also
theoretically\cite{wu_prb, theory2, wu_prb2} for different materials
containing similar environments for a Co$^{3+}$ ion. Even for
GdBaCo$_2$O$_{5.5}$, calculations\cite{hf_prb,hf_gdbacoo} and theoretical
results\cite{khalyavin} show a HS state as the most
stable for Co$_{pyr}$, contrary to our results.
However, all the solutions we tried converge to an IS state of Co$_{pyr}$
indicating an enormous stability of such a state. 
The different results obtained with similar methods could be due to our
structure optimization, that leads to changes of up to 2\% in interatomic
distances and bond angles with respect to the experimental room
temperature data. It is well-known that variations in the geometry of the
cation environment could lead to the stabilization of a different spin
state.\cite{novak} Also, the LDA+U method is strongly dependent on the
starting point and several solutions can be obtained depending on the
starting density chosen.
On the other hand, for Co$_{oct}$, 
different spin states could be converged: a LS state, an IS state and a
spin state order with half the Co$_{oct}$ atoms in a HS state and the
other half in a LS state.
A HS state of Co$_{oct}$ could
not be converged with the structural data we utilized. It has been shown
experimentally that a structural transition\cite{frontera_prb} occurs at
the metal-insulator transition and this has been related to the appearance
of a HS state. 
Such a HS state would stabilize through
a strong spin-lattice coupling that our zero temperature DFT calculations
cannot describe.

\subsection{Magnetic ordering and magnetic transitions}\label{fursec}

\begin{figure}
\includegraphics[width=0.8\columnwidth]{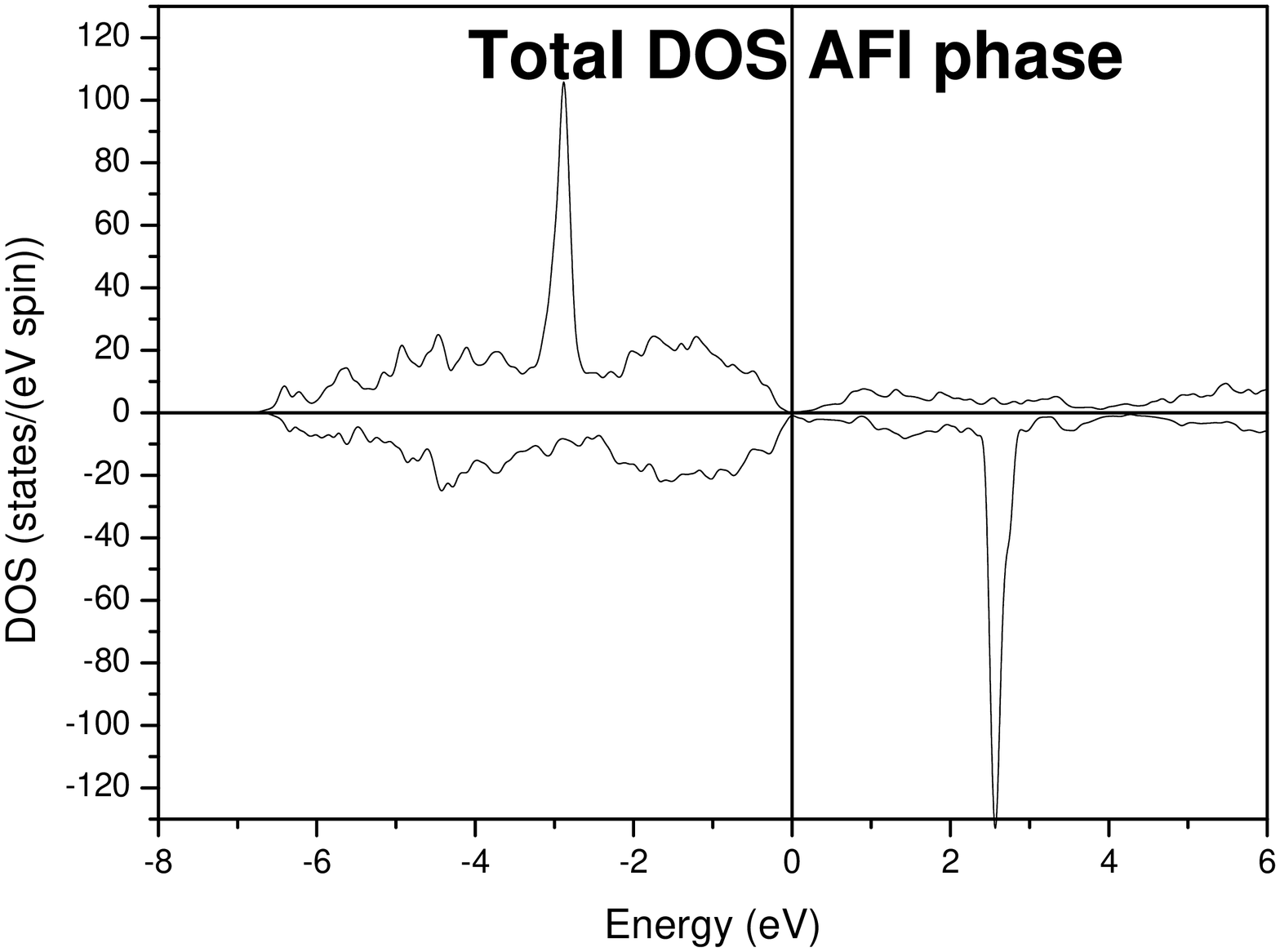}
\includegraphics[width=0.8\columnwidth]{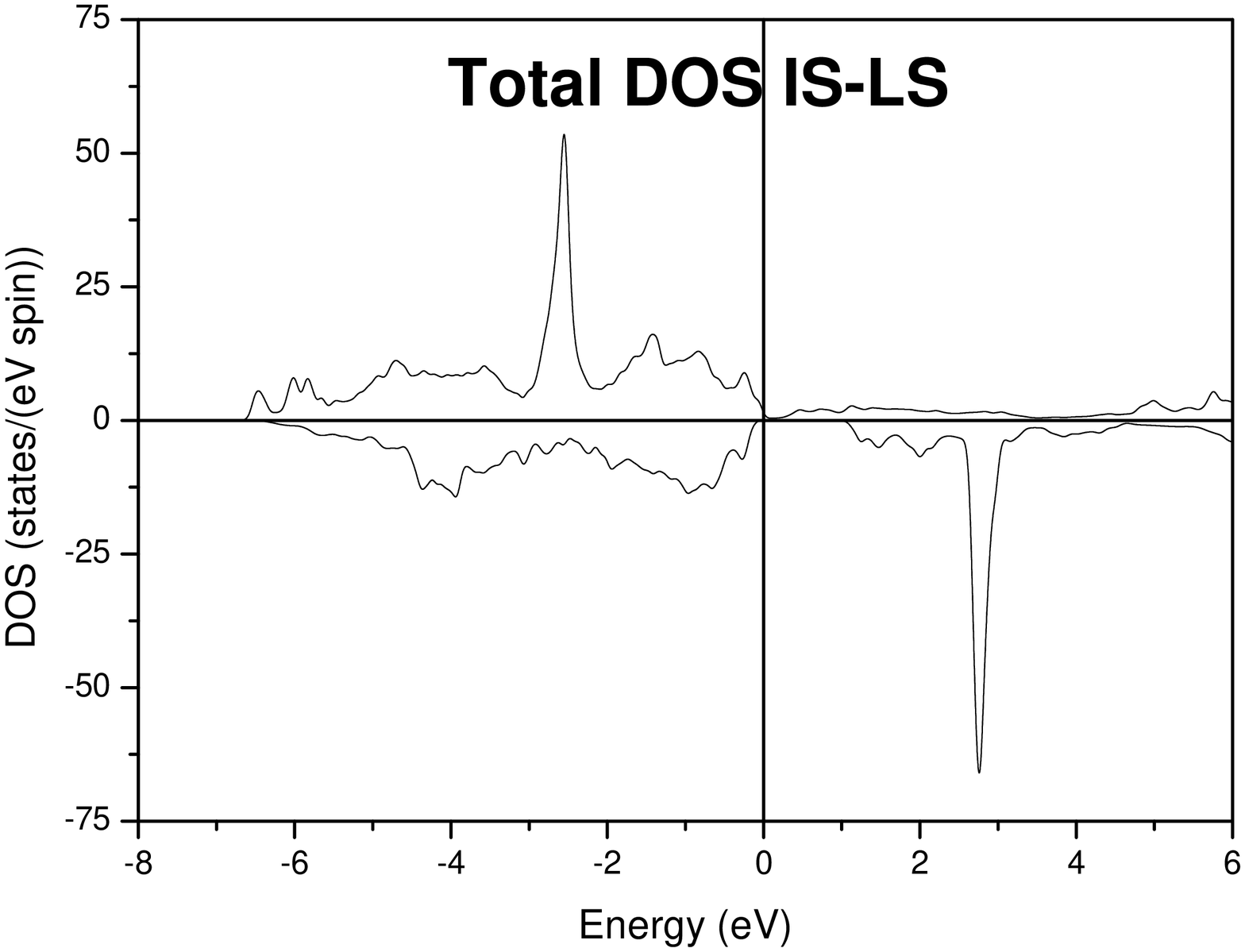}
\caption{Spin-up and spin-down total DOS plots for the material in the
insulating phases, when the Co$_{oct}$ atoms are in a LS state. The upper
panel shows an AF coupling between Co$_{pyr}$ layers and the lower panel
shows a FM situation.}\label{afi_fmi}
\end{figure}

We have studied several magnetic and spin configurations within the LDA+U
approximation. Both FM and AF couplings along the three crystallographic
directions have been
considered. Our calculations yield a FM coupling along the b- and $c$-axis as the
most stable one for any spin state of the Co$_{oct}$ layer.
Within each of the layers, an AF ordering
cannot be stabilized. It has been argued\cite{magnprops} that orbital
ordering produces the stabilization of a FM ordering within the plane, and
that is what we observe (see Section \ref{orbsec}). For
the case of a LS state of the Co$_{oct}$ atoms, a FM and an AF ordering
of Co$_{pyr}$ layers mediated by a nonmagnetic Co$_{oct}$ layer were calculated. AF
ordering is more stable by some 10 meV/Co (U-dependent value but
consistent for a wide set of U choices), not a strong magnetic coupling.
Both IS-LS solutions (IS stands for the spin state
of Co$_{pyr}$ and LS for the spin state of the Co$_{oct}$ atoms), consisting of planes coupled AF or FM, lead to an insulating
behavior (see Fig. \ref{afi_fmi}). We observe that the IS-LS configuration
leads to a zero-gap insulator, in accordance
with experiments observing a narrow gap.\cite{taskin} This is the only
solution that could explain the properties of the AFI (antiferromagnetic
insulator) phase found below 70 K. The weak magnetic coupling between the
layers breaks up at that temperature entering the paramagnetic phase above
it.

From our calculations, we interpret the AF-FM transition that occurs at
about 220 K as a change in the magnetic coupling of FM Co$_{pyr}$ layers
mediated by nonmagnetic Co$_{oct}$ planes, because such a FM ordering is
the only one leading to a FMI (ferromagnetic insulator) phase. This is consistent with the
experimental observations in Ref. \onlinecite{magnprops}.

\begin{figure}
\includegraphics[width=0.8\columnwidth]{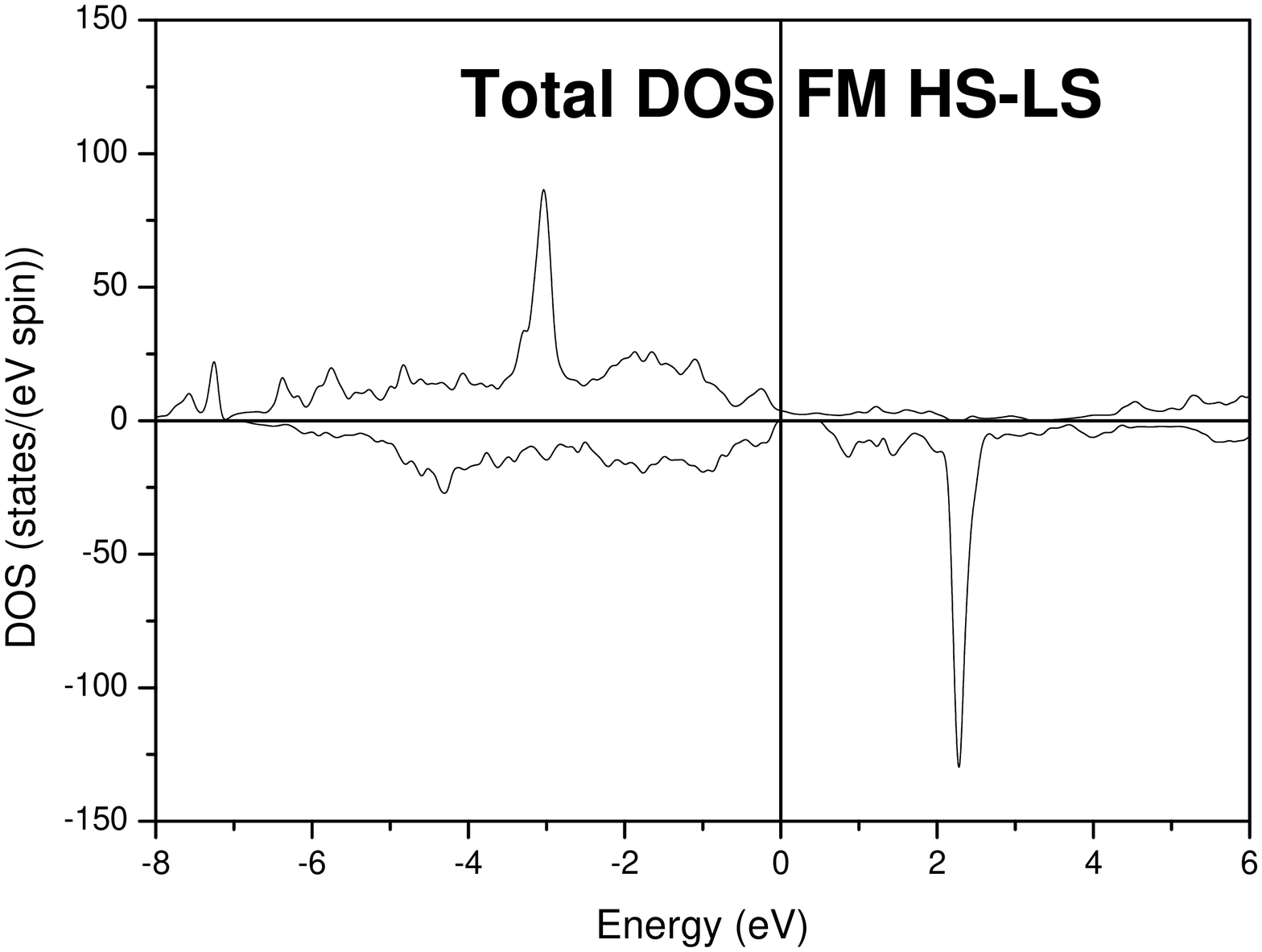}
\includegraphics[width=0.8\columnwidth]{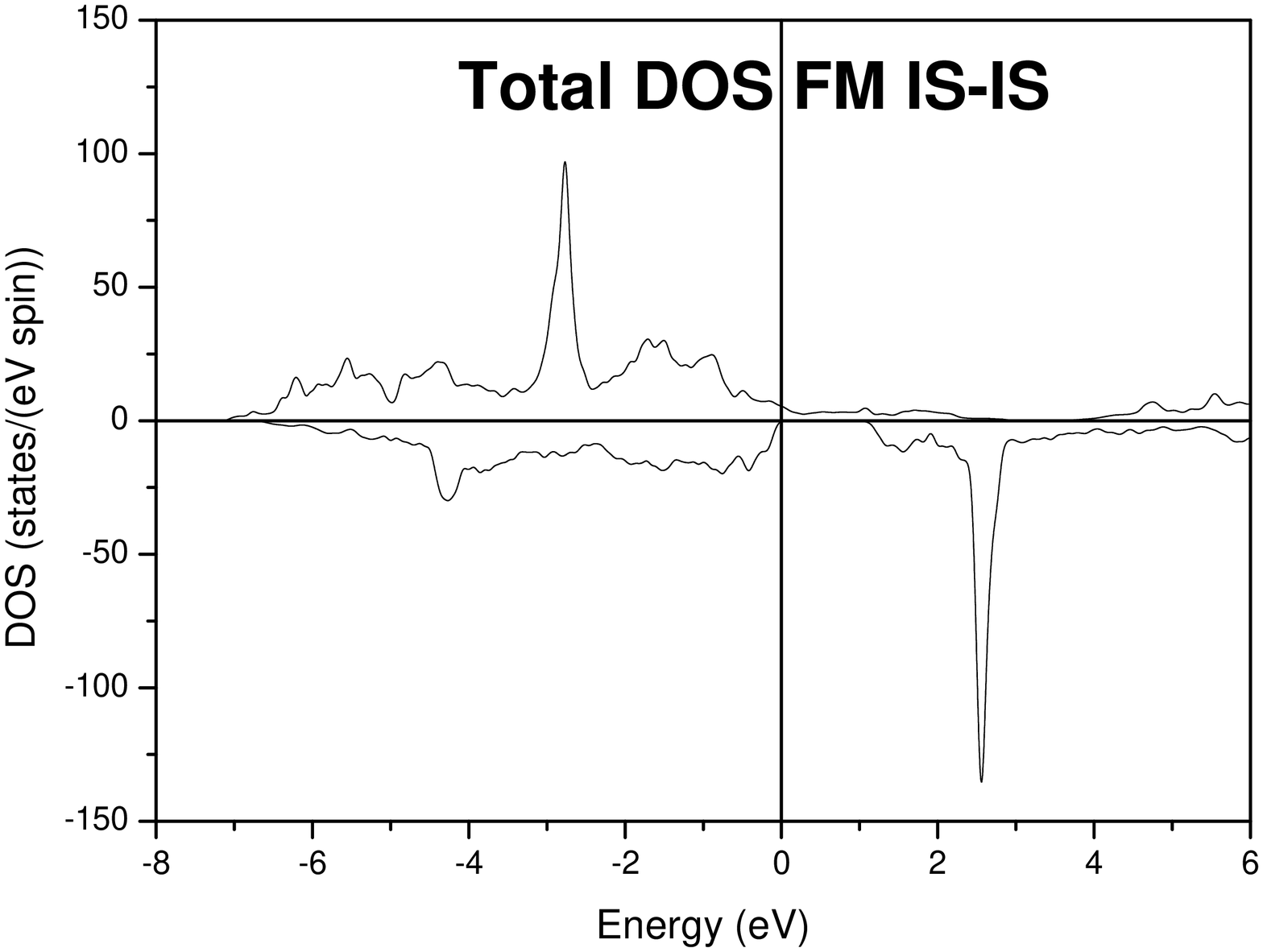}
\caption{Spin-up and spin-down total DOS plots for the material in the
possible metallic phases
considered in the text. The upper panel shows a spin state ordering HS-LS in
the Co$_{oct}$ layers and the lower panel shows an IS state for all the
Co$_{oct}$ atoms.}\label{metals}
\end{figure}

The magnetic coupling
between Co$_{oct}$ and Co$_{pyr}$ layers when the former acquires an IS
state is FM. The energy difference between a FM and an AF configuration is
about 40 meV/Co (U-dependent value), a strong magnetic coupling.
We could also converge a FM solution where half of the
Co$_{oct}$ atoms are in a HS state and the other half in a LS state, a
spin state ordering situation, similar to what has been observed for
NdBaCo$_2$O$_{5.5}$\cite{fauth} (in this case it has been predicted to be
between IS and LS states) or predicted
theoretically for the whole series\cite{khalyavin} (but in that case the
authors assume that Co$_{pyr}$ posses a HS). These two solutions we have
computed lead to a
half-metallic state (see Fig. \ref{metals}). In principle, any of these could  be the
magnetic ordering in the metallic phase above the metal-insulator
transition. But, since a structural deformation has been observed
experimentally at the metal-insulator transition\cite{frontera_prb} and this has
been ascribed to the onset of a HS state, we should not rule out the
possibility of a HS Co$_{oct}$ layer above the metal-insulator transition.
The structural data we utilized (calculated at T= 0) does not allow to
stabilize that configuration. The calculation of such a phase by DFT
techniques would require a precise knowledge of the geometry above 360 K.
The recent magnetization measurements\cite{magnprops} observing
a total magnetic moment of 1 $\mu_B$/Co above the metal-insulator
transition would rule out this possibility. As we will see below, orbital
angular momenta need to be taken into account for explaining the total
magnetization measurements.

\subsection{Orbital ordering}\label{orbsec}

Co$^{3+}$ ions in an IS state have an orbital 
degree of freedom, that could lead to
the existence of orbital ordering along the
b and c-axes.\cite{taskin,goodenough_jssc}
The octahedral environment is elongated along the crystallographic c-axis (we will consider it
here as our z-axis for the representation of the $d$-levels). For the
pyramidal environment, the z-axis will be considered along the direction
of the oxygen vacant (crystallographic b-axis). For Co$_{pyr}$, due to the
small crystal field splitting a Co$^{3+}$ ion suffers, an orbital
degeneracy is likely to occur as well (depending on the intra-atomic
exchange energy).

\begin{figure}
\includegraphics[width=0.8\columnwidth]{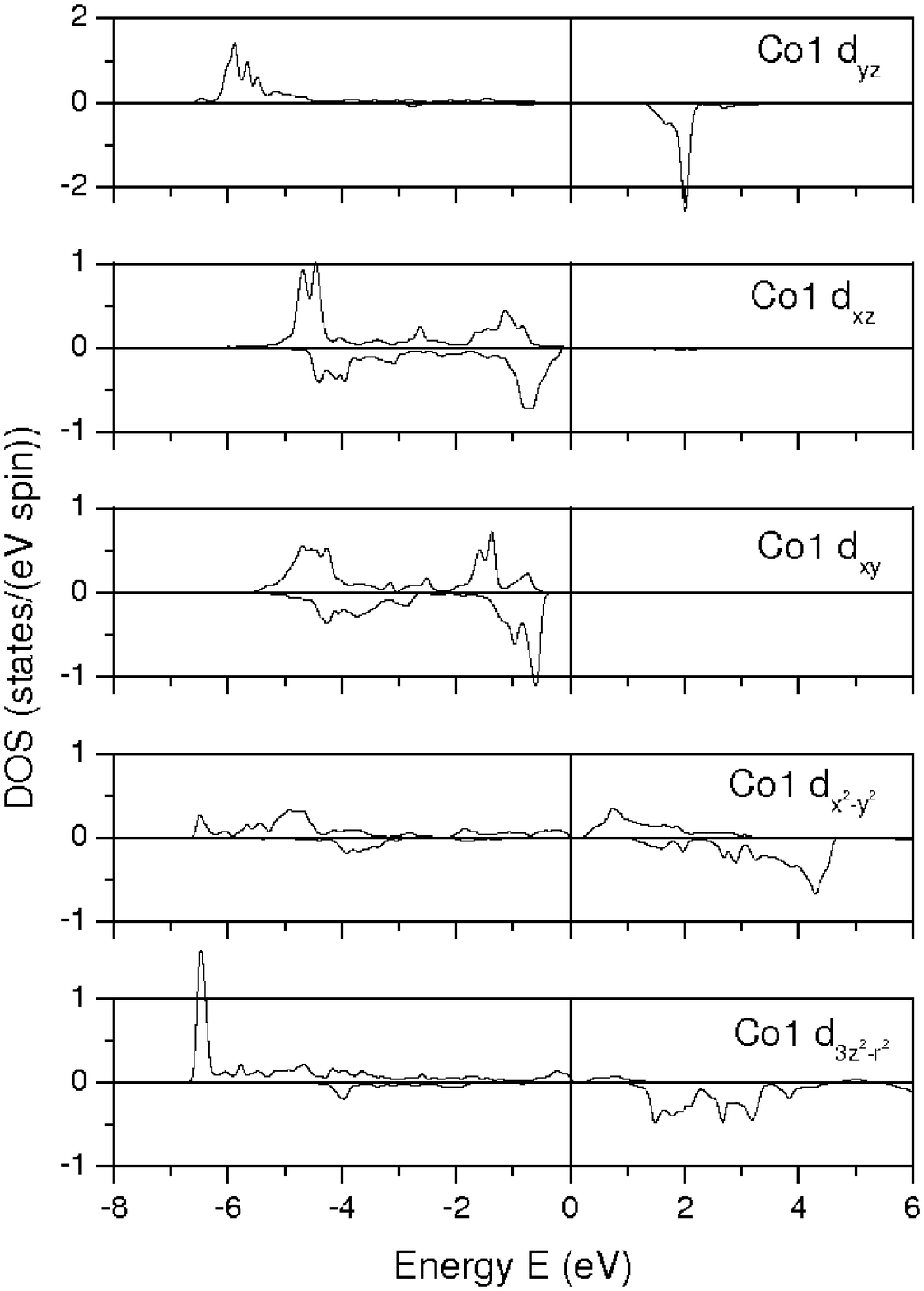}
\caption{Spin-up and spin-down partial density of states of Co1 (in a
square pyramidal environment). It
can be observed that it is in an IS state and the hole in the t$_{2g}$
multiplet is in the $d_{yz}$ orbital.}\label{dosco1pyr}
\end{figure}

\begin{figure}
\includegraphics[width=0.8\columnwidth]{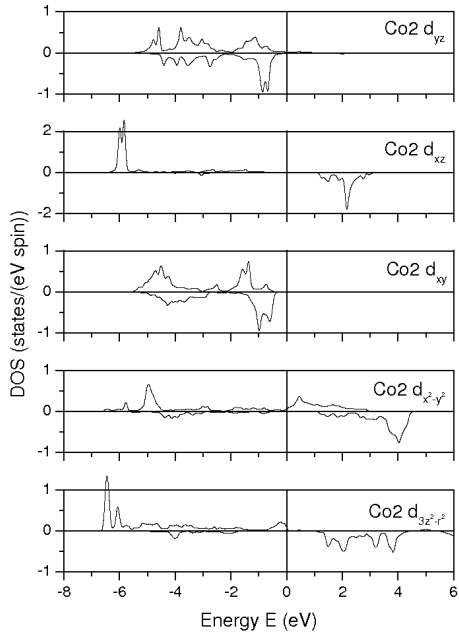}
\caption{Spin-up and spin-down partial density of states of Co2 (in a
square pyramidal environment). It
can be observed that it is in an IS state and the hole in the t$_{2g}$
multiplet is in the $d_{xz}$ orbital.}\label{dosco2pyr}
\end{figure}

Let us first begin with the IS-LS situation, the ground state. Figs. \ref{dosco1pyr} and
\ref{dosco2pyr} show the density of states (DOS) of the $d$-levels for the IS state of the Co$_{pyr}$ atoms.
Their electronic structure does not vary much when the spin state of
Co$_{oct}$ changes. According to our calculations,
the IS state of Co$_{pyr}$ is fairly stable and
remains unchanged when increasing temperature, so our description of this
state is valid for any spin state of the Co$_{oct}$ atoms. We can
observe in the curves a different electronic structure in the
crystallographically equivalent Co atoms,
indicating the existence of orbital ordering along the $b$-axis (the
crystallographic axis that joins the two inequivalent Co$_{pyr}$ (and
also Co$_{oct}$)
atoms we have considered in our
calculations (see Fig. \ref{strfig}). In Ref.
\onlinecite{goodenough_jssc}, this orbital state is predicted to be
$d_{xz}$ $\pm$ $i$ $d_{yz}$ for DyBaCo$_2$O$_{5.5}$, but we obtain as the ground state an orbital ordering
between real orbitals ($d_{xz}$ and $d_{yz}$), being this configuration some 20 meV/Co
more stable than the state with equal orbital structure along the $b$-axis
(even though the magnitude of the energy difference is dependent on the value of U chosen, the
important point is that it certainly is more stable independently of U), a
value comparable to the energy involved in magnetic ordering. Similar
calculations indicate that the same orbital ordering is stable as well
along the c-axis. Couplings within the bc-plane are equivalent both
orbitally and magnetically.
However, the
location of the electron that is promoted into the e$_g$ doublet is more
difficult to ascribe, since it is not in a pure $d_{3z^2-r^2}$ nor in the
$d_{x^2-y^2}$ orbital, as can be observed in Figs. \ref{dosco1pyr} and
\ref{dosco2pyr}. We also
decomposed the DOS curves in a set of orbitals using the complex functions
d$_{3z^2-r^2}$ $\pm$ $i$ d$_{x^2-y^2}$ (consistent with having the same
environment for both Co atoms, since these orbitals have cubic symmetry) but the electron does not fit into those
orbitals alternately neither. Some more complicated orbital structure occurs that
we could not discern with our calculations. We can only ascertain that the
resultant states must have some component of each orbital, probably a
complex combination since an unquenched orbital angular momentum occurs as
we will see below. Also, the
e$_{g}$ bands (this is valid for Co$_{oct}$ in an IS state as well)
contain more than one electron, that probably comes from a very strong
hybridization of the e$_{g}$ band with the surrounding
O $p$ bands. This could imply the formation of a Co$^{2+}$\underline{L}
state (\underline{L} being a ligand hole). This strong hybridization had
been already found in previous works,\cite{flavell_prb,hf_gdbacoo} but for
a HS state of Co$_{pyr}$.

\begin{figure}
\includegraphics[width=0.8\columnwidth]{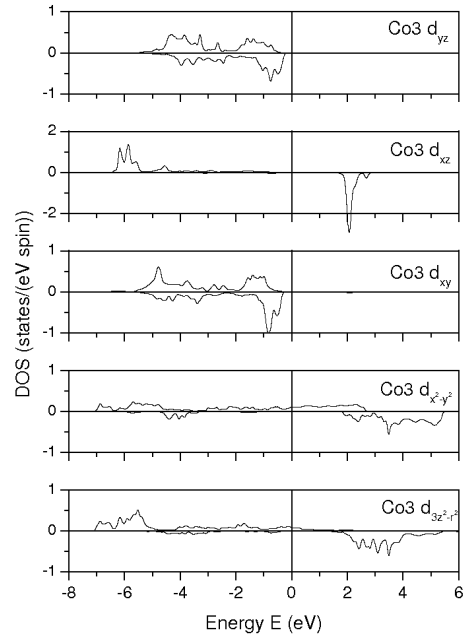}
\caption{Spin-up and spin-down partial density of states of Co3 (in an
octahedral environment) for
the IS-IS configuration. It
can be observed that it is in an IS state and the hole in the t$_{2g}$
multiplet is in the $d_{xz}$ orbital.}\label{dosco3oct}
\end{figure}

\begin{figure}
\includegraphics[width=0.8\columnwidth]{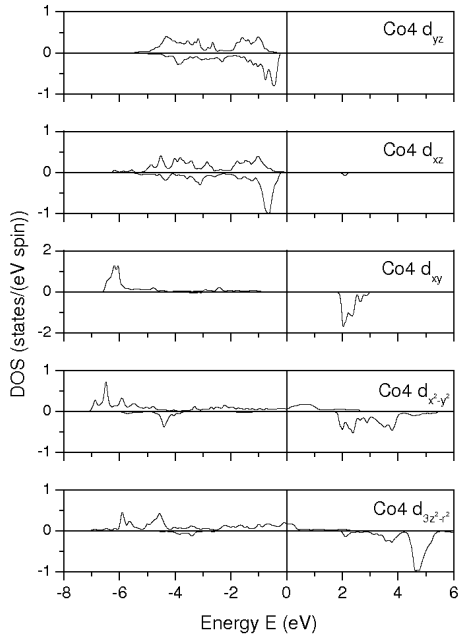}
\caption{Spin-up and spin-down partial density of states of Co4 (in an
octahedral environment) for
the IS-IS configuration. It
can be observed that it is in an IS state and the hole in the t$_{2g}$
multiplet is in the $d_{xy}$ orbital.}\label{dosco4oct}
\end{figure}

For the case of Co$_{oct}$, we present in Figs. \ref{dosco3oct} and
\ref{dosco4oct} the $d$-DOS for the IS state. The plots of the LS state are
not introduced 
because the electronic structure is basically a
t$_{2g}^6$ e$_g^0$ configuration in a nearly octahedral environment, the
t$_{2g}$ levels fully occupied and the e$_{g}$ levels fully unoccupied.
In the IS state, one of the
electrons in the lower lying multiplet is promoted to the e$_{g}$ levels.
There occurs an orbital ordering phenomenon as well for Co$_{oct}$ in the
IS state along the $b$-axis. The DOS of two contiguous Co$_{oct}$ atoms along that
direction are shown in Figs. \ref{dosco3oct} and \ref{dosco4oct}, where it
can be seen that the hole left in the t$_{2g}$ multiplet is alternatively at the
$d_{xy}$ and $d_{xz}$ orbitals. This configuration is some 10 meV/Co
more stable than the non-orbitally ordered scenario (that energy value
might vary slightly with the value of U chosen, but it is consistent for a
wide range of U).
The electron that is
promoted in the e$_{g}$ multiplet places in alternate orbitals,
but from our calculations it is difficult to describe the
combination of $d_{3z^2-r^2}$ and $d_{x^2-y^2}$ orbitals that forms the doublet. We
know it is not the pure real orbitals as we can see in Figs. \ref{dosco3oct} and
\ref{dosco4oct} but they are not the complex orbitals $d_{3z^2-r^2}$ $\pm$ $i$
$d_{x^2-y^2}$ neither.
We tried that possibility decomposing our DOS in a
basis containing those two orbitals but such an ordering was not found.  Some other admixture of these orbitals is happening due to the
local environment of the atom but from our calculations this could not be
sorted out. We encounter here a similar situation to the case
of Co$_{pyr}$, 
the fact that there exist some unquenched orbital angular momenta rules out
the possibility of having real orbitals, the electron must be located in 
a state formed by some complex
combination of orbitals. The e$_{g}$ band is also in this case strongly
hybridized, possibly forming a Co$^{2+}$\underline{L} state as for Co$_{pyr}$.  
In an IS-IS situation, the orbital ordering was found to be
stable both along the b- and c-axis, consistent with the picture
established in Ref.
\onlinecite{magnprops}.

In a spin state ordering situation, the Co$_{oct}$ atom in a LS state
acquires some magnetic moment due to the surrounding HS ions and the e$_g$
bands become occupied forming a very wide band that crosses the Fermi
level leading to the half-metallic behavior observed in Fig. \ref{metals}.

\subsection{Spin-orbit study: unquenched orbital angular
momenta and magnetic anisotropy}\label{sosec}

We have also carried out calculations including spin-orbit coupling to
elucidate how the magnetic anisotropy changes for the different spin
states and magnetic couplings and to evaluate the orbital angular momenta
(OAM) of
the different Co atoms in the structure, that had been predicted to have
some unquenched OAM\cite{goodenough_jssc} but whose values have never been
measured. Only values of total magnetization are available in the
literature,\cite{kim,respaud_prb,roy} showing some
discrepancies between them.

\begin{table}[h]
\caption{Values of the orbital angular momenta of the four Co atoms in the
structure in the three spin state configurations studied. The values are
given in $\mu_B$ and are considered inside the muffin-tin
sphere. These values are calculated for the magnetization along the
experimental easy
axis.}\label{taboam}
\begin{tabular}{|c|c|c|c|}
\hline
Atom & IS-IS ($\mu_B$) & IS-LS ($\mu_B$) & IS-(HS-LS)($\mu_B$)\\
\hline
\hline
Co1 (pyr) & 0.30 & 0.28 & 0.27\\
Co2 (pyr) & 0.02 & 0.01 & 0.01\\
Co3 (oct) & 0.01 & -0.05 & 0.03 (LS)\\
Co4 (oct) & 0.24 & -0.07 & 0.19 (HS)\\
\hline
\end{tabular}
\end{table}

A summary of the OAM calculated is presented in
Table \ref{taboam}. These data are 
calculated assuming the magnetization goes along the experimental easy
axis ($c$-axis). The values are given
inside the muffin-tin spheres, so the real values of the moments are, hence,
somewhat bigger (about 10-30\%, but this is difficult to estimate).  
They also are U-dependent by up to 25\% for other
combinations of U parameters within a reasonable range. Large values up to
0.6 $\mu_B$ for some of the Co atoms would be consistent with
our calculations. Experimental measurements are needed to confirm this
observation. We performed these calculations for the three different spin
state configurations we could converge for the Co$_{oct}$ atoms
(Co$_{pyr}$ is always in an IS state): a LS, an IS state and a spin state
ordering with half the Co$_{oct}$ atoms in a HS state and the other half
in a LS state.

Irrespective of the Co$_{oct}$ spin state, we observe that
Co$_{pyr}$ atoms have very similar
angular momenta (one of them being about 0.3 $\mu_B$ inside the muffin-tin sphere). The
electronic structure of these atoms barely
changes when a spin state transition occurs to Co$_{oct}$. This atom
develops a big angular momentum in one of the sites when it is in an IS
state. It is worth noting that the big orbital angular momentum occurs
when the hole in the t$_{2g}$ multiplet is left in the $d_{xz}$ orbital,
both in the case of Co$_{pyr}$ and Co$_{oct}$ in an IS state, whereas the
configuration with a hole in a different t$_{2g}$ orbital produces a
negligible moment. Small values are predicted also for Co$_{oct}$ in a LS
sate. These are 
 oppositely oriented to the magnetic moment of the Co$_{pyr}$ ions when
all the Co$_{oct}$ atoms are in a LS state and parallel to them when the
layer includes Co$_{oct}$ atoms in a HS state. These also have a large
value of the OAM, but smaller than those of the IS state.

We also performed calculations of the magnetic anisotropy energies
and our results agree with the experimental data
available.\cite{taskin}
We obtain a strong Ising-like behavior for the AFI phase (IS-LS with IS
Co$_{pyr}$ planes coupled AF via nonmagnetic Co$_{oct}$ layers). In that
case, the c-axis is the easy axis by some 150 K/Co,
a big anisotropy energy, in agreement with the experimental
value (some 80-100
K/Co).\cite{taskin}
For the IS-LS FMI phase, we find an easy axis along the
c-axis, being the b-direction almost degenerate (3 K/Co harder) and the $a$-direction harder by some 40 K/Co. No trace of
Ising-like behavior is found for this phase. This disagrees with the
experimental findings because our zero-temperature DFT calculations
cannot describe appropriately the magnetic anisotropy properties at 260 K,
where they have been measured.\cite{taskin} A similar situation occurs
for the IS-IS FM state, the b-axis is
the easy axis being the a and c directions harder by some 30 K/Co. In
these cases it could be possible that the domain formation found in Ref.
\onlinecite{magnprops} could have an influence on the value of the
magnetic anisotropy constants that is not considered in the calculation of
the magnetocrystalline anisotropy energy.

\section{Conclusions}

In this paper we have presented a study of the electronic structure and magnetic
properties of the Co oxide GdBaCo$_2$O$_{5.5}$ using \emph{ab
initio} calculations considering an all-electron, full-potential APW+lo
method. The points we tried to address were the spin state
transitions, the orbital ordering and spin-orbit effects (unquenched
orbital angular momenta and magnetic anisotropy properties).
From our calculations we confirm the stability of the IS state for Co$_{pyr}$ or the
conduction properties of the IS-LS state as a narrow gap insulator. The
transition to a metallic phase must be accompanied of a spin state
transition of Co$_{oct}$. It is necessary to know the precise geometry in 
the metallic phase to determine its spin and magnetic configuration.
We have observed the material is in an orbitally-ordered state both at low
temperatures in the IS-LS state and also in a hypothetical IS-IS state, and we described the
orbitals involved in the phenomenon. We also predict big unquenched orbital
angular momenta for some of the Co atoms in the structure, 
a fact that had
been predicted but whose values have never been measured. We estimate
their values and explain their origin. Also, the magnetic anisotropy was
studied and agrees with the experimental observation of a strong
Ising-like behavior. In summary, we
establish a consistent picture for the properties of the material by
confirming some experimental evidences and predicting new results based on
first principles calculations.

\begin{acknowledgments}
The authors wish to thank the CESGA (Centro de Supercomputacion de
Galicia) for the computing facilities and the
Xunta de Galicia for the financial support through a grant and the project
PGIDIT02TMT20601PR.
\end{acknowledgments}


\end{document}